\documentclass[aps, prb, twocolumn, superscriptaddress, amsmath,
amssymb, showpacs]{revtex4}

\usepackage{graphicx}
\usepackage[pdftex]{hyperref}

\begin{document}

\title{Discontinuous Euler instability in nanoelectromechanical systems}

\author{Guillaume Weick}
\affiliation{Dahlem Center for Complex Quantum Systems and Fachbereich Physik, Freie Universit\"at Berlin,
D-14195 Berlin, Germany}
\affiliation{IPCMS (UMR 7504), CNRS and Universit\'e de Strasbourg, F-67034 Strasbourg, France}
\author{Fabio Pistolesi}
\affiliation{CPMOH (UMR 5798), CNRS and Universit\'e de Bordeaux I, F-33405 Talence, France}
\affiliation{LPMMC (UMR 5493),
CNRS and Universit\'e Joseph Fourier, F-38042 Grenoble, France}
\author{Eros Mariani}
\affiliation{Dahlem Center for Complex Quantum Systems and Fachbereich Physik, Freie Universit\"at Berlin,
D-14195 Berlin, Germany}
\affiliation{School of Physics, University of Exeter, Stocker Road, Exeter, EX4
4QL, UK}
\author{Felix von Oppen}
\affiliation{Dahlem Center for Complex Quantum Systems and Fachbereich Physik, Freie Universit\"at Berlin,
D-14195 Berlin, Germany}

\date{\today}

\begin{abstract}
We investigate nanoelectromechanical systems near mechanical instabilities. We
show that quite generally, the interaction between the electronic and the
vibronic degrees of freedom can be accounted for essentially exactly when the
instability is continuous. We apply our general framework to the Euler buckling
instability and find that the interaction between electronic and vibronic
degrees of freedom qualitatively affects the mechanical instability, turning it
into a discontinuous one in close analogy with tricritical points in the Landau theory of phase transitions. 
\end{abstract}


\maketitle

\section{Introdution}
\begin{figure}[t]
\includegraphics[width=\columnwidth]{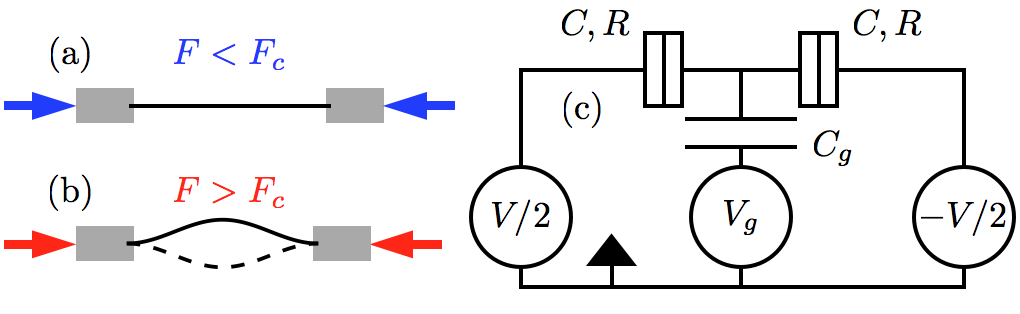}
\caption{\label{fig:buckled_CNT}%
(color online) Sketch of a nanobeam
(a) in the flat state and (b) the buckled state with two equivalent metastable positions of the rod (solid and dashed lines). An equivalent circuit of the embedded SET is shown in (c).}
\end{figure}

The buckling of an elastic rod by a longitudinal
compression force $F$ applied to its two ends constitutes the paradigm of a
mechanical instability, called buckling instability.\cite{landau} It was first
studied by Euler in 1744 while investigating the maximal load that a column can
sustain.\cite{euler} As long as $F$ stays below a critical force $F_c$, the rod
remains straight, while for $F>F_c$ it buckles, as sketched in
Fig.~\ref{fig:buckled_CNT}a-b. The transition between the two states is continuous and the frequency of the fundamental bending mode vanishes at the instability.

There has been much recent interest in exploring buckling instabilities in
nanomechanical systems. In the quest to understand the remarkable mechanical
properties of nanotubes,\cite{ponch99_Science, huettel_Science09,
bachtold_Science09} there have been observations of compressive buckling
instabilities in this system.\cite{falvo97_Nature} The Euler buckling
instability has been observed in SiO$_2$ nanobeams and shown to obey continuum
elasticity theory.\cite{carr03_APL} There are also close relations with the
recently observed wrinkling \cite{lau09} and possibly with the rippling
\cite{meyer07_Nature} of suspended graphene samples. Theoretical works have
studied the quantum properties of nanobeams near the Euler instability,
\cite{carr01_PRB, werne04_EPL, peano06_NJP, savel06_NJP} proposing this system
to explore zero-point fluctuations of a mechanical mode \cite{werne04_EPL} or to
serve as a mechanical qubit. \cite{savel06_NJP} 

In this work, we study the interaction of current flow with the vibrational motion near such continuous mechanical instabilities which constitutes a fundamental issue of nanoelectromechanics. \cite{craig00_Science} Remarkably, we find that under quite general conditions, this problem admits an essentially exact solution due to the continuity of the instability and the consequent vanishing of the vibronic frequency at the transition (``critical slowing down"). In fact, the vanishing of the frequency implies that the mechanical motion becomes slow compared to the electronic dynamics and an appropriate non-equilibrium Born-Oppenheimer (NEBO) approximation becomes asymp\-totically exact near the transition. Here, we illustrate our general framework by applying it to the nanoelectromechanics of the Euler instability.

We find that the interplay of electronic transport and the mechanical instability causes significant qualitative changes both in the nature of the buckling and in the transport properties. In leading order, the NEBO approximation yields a current-induced conservative force acting on the vibronic mode. At this order, our principal conclusion is that the coupling to the electronic dynamics can change the nature of the buckling instability from a continuous to a discontinuous transition which is closely analogous to tricritical behavior in the Landau theory of phase transitions. Including in addition the fluctuations of the current-induced force as well as the corresponding dissipation leads to Langevin dynamics of the vibrational mode which becomes important in the vicinity of the discontinuous transition. Employing the same NEBO limit to deduce the electronic current, we find that the buckling instability induces a current blockade over a wide range of parameters. This is a manifestation of the Franck-Condon blockade \cite{koch05_PRL, letur09_NaturePhysics, pisto07_PRB} whenever the buckling instability remains continuous but is caused by a novel tricritical blockade when the instability is discontinuous. The emergence of a current blockade in the buckled state suggests that our setup could, in principle, serve as a mechanically-controlled switching device.

\section{Model}
Close to the Euler instability, the frequency of the
fundamental bending mode of the beam approaches zero, while all higher modes
have a finite frequency. \cite{landau} This allows us to retain only the
fundamental mode of amplitude $X$ (see Fig.~\ref{fig:buckled_CNT}a-b) and
following previous studies, \cite{carr01_PRB, werne04_EPL, peano06_NJP} we reduce the vibrational Hamiltonian to \cite{footnote_zeromode}
\begin{equation}
\label{eq:H_vib}
H_\textrm{vib}=\frac{P^2}{2m}+\frac{m\omega^2}{2}X^2+\frac{\alpha}{4}X^4,
\end{equation}
which is closely analogous to the Landau theory of continuous phase transitions. In Eq.~(\ref{eq:H_vib}), $P$ is the momentum conjugate to $X$ and $m$ denotes an effective mass. The mode frequency $\omega^2 \sim 1-F/F_c$ changes sign when $F$ reaches a critical force $F_c$. Global stability then requires a quartic term with $\alpha>0$. Thus, for $F<F_c$ ($\omega^2>0$), $X=0$ is the only stable minimum and the beam remains straight. For $F>F_c$ ($\omega^2<0$), the beam buckles into one of the two minima at $\pm X_+=\pm\sqrt{-m\omega^2/\alpha}$. 

\begin{figure*}[tbh]
\includegraphics[width=2\columnwidth]{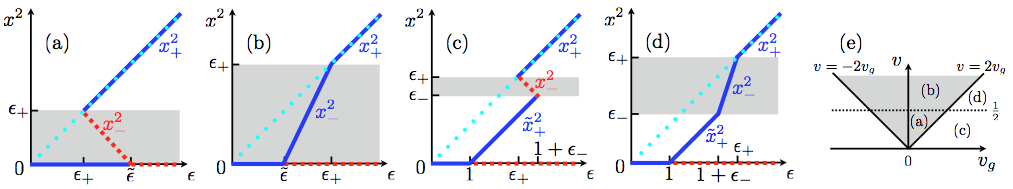}
\caption{\label{fig:stable_solutions}%
(color online) (Meta)stable (solid blue lines) and unstable (dashed red lines) positions of the nanobeam vs.\ scaled force $\epsilon$ for 
(a) $|v_g|<v/2$, $v<1/2$, 
(b) $|v_g|<v/2$, $v>1/2$, 
(c) $v_g>v/2$, $v<1/2$ (for $\epsilon_+>1$; a similar plot holds for $\epsilon_+<1$),
(d) $v_g>v/2$, $v>1/2$, as indicated in the $v_g$--$v$ plane in (e). The dotted blue line is the result without electron-vibron coupling. Notation: $\epsilon_\pm=2v_g\pm v$, $\tilde\epsilon=1/2+v_g/v$, $x_+^2=\epsilon$, $\tilde x_+^2=\epsilon-1$, and $x_-^2=(\epsilon-\tilde\epsilon)/(1-1/2v)$. Grey indicates conducting regions.}
\end{figure*}

The vibronic mode of the nanobeam interacts with an embedded metallic
single-electron transistor (SET), consisting of a small metallic island coupled
to source, drain, and gate electrode (Fig.~\ref{fig:buckled_CNT}c). We assume that the SET operates in the Coulomb blockade regime \cite{dittrich} and that bias and gate voltage are tuned to the vicinity of the conducting region between SET states with, say, zero and one excess electron. The electronic degrees of freedom couple to the vibronic motion through the occupation $\hat n$ of excess electrons on the metallic island. Specifically, we assume that the electron-vibron coupling does not break the underlying parity symmetry of the vibronic dynamics under $X\to -X$. This follows naturally when the coupling emerges from the electron-phonon coupling {\em intrinsic} to the nanobeam \cite{maria09_preprint} and implies that the coupling depends only on even powers of the vibronic mode coordinate $X$. The dominant coupling is quadratic in $X$,  
\begin{equation}
\label{eq:H_c}
H_c=\frac{g}{2}X^2\hat n, 
\end{equation}
with a coupling constant $g>0$. \cite{maria09_preprint} When there is a significant contribution to the electron-vibron coupling originating from the electrostatic dot-gate interaction, we envision a symmetric gate setup consistent with Eq.~(\ref{eq:H_c}).

In the presence of the vibronic dynamics $X(t)$, the electronic occupation $n(X,t)$ of the island is described by the Boltzmann-Langevin equation \cite{blanter}
\begin{equation}
  \frac{dn}{dt} = \{n, H_{\rm vib}\} + \Gamma_+ (1 - n) - \Gamma_- n + \delta J_+ - \delta J_-.
  \label{BL}
\end{equation}
This equation assumes that the bias is large compared to temperature so that
tunneling is effectively unidirectional and the relevant tunneling rates
$\Gamma_\pm$ for tunneling onto ($+$) and off ($-$) the island are given by
$\Gamma_\pm=R^{-1}(V/2\pm\bar V_g)\Theta(V/2\pm\bar V_g)$. Here, $R$ denotes the
tunneling resistances ($R\gg h/e^2$) between island and leads, $V$ is the bias
voltage, $\Theta(x)$ denotes the Heaviside step function, and we set
$\hbar=e=1$. Since both gate voltage (via the capacitances $C$ and $C_g$ in
Fig.~\ref{fig:buckled_CNT}c) and vibronic deformations couple to the excess
charge $\hat n$ on the island, the effective gate voltage $\bar V_g = V_g -
gX^2/2$ combines the gate potential $V_g$ (measured from the degeneracy point between the states with zero and one excess electron) and the vibron-induced shift of the electronic energy described by $H_c$. The stochastic Poisson nature of electronic tunneling is accounted for by including the Langevin sources $\delta J_\pm$ with correlators $\langle \delta J_+(t)\delta J_+(t')\rangle = \Gamma_+ (1-n)\delta(t-t')$ and $\langle \delta J_-(t)\delta J_-(t')\rangle = \Gamma_- n\delta(t-t')$. The vibronic dynamics enters Eq.\ (\ref{BL}) through the Poisson bracket $\{n, H_{\rm vib}\}$.

\section{Stability analysis}
We are now in a position to investigate the
influence of the electronic dynamics on the vibronic motion. Near the
instability, the vibrational dynamics becomes slow compared to the electronic
tunneling dynamics. As has recently been shown, \cite{blant04_PRL,marti06_PRB} the effect of the current on the vibrational motion can then be described within a NEBO approximation in which the vibrational motion is subject to a current-induced force $-g X n(X,t)$ originating in the electron-vibron interaction (\ref{eq:H_c}). 
This current-induced force involves both a time-averaged and conservative force as well as fluctuating and frictional forces, resulting in Langevin dynamics of the vibronic degree of freedom.

In lowest order, the Langevin dynamics only involves the conservative force which emerges from the average occupation 
$n_0(X)$ in the absence of fluctuations ($\delta J_\pm \simeq 0$) and vibronic dynamics ($\{n, H_{\rm vib}\}\simeq 0$). In this limit, Eq.\ (\ref{BL}) reduces to the usual rate equation of a metallic SET so that \cite{dittrich}
\begin{equation}
\label{eq:n(x)}
n_0(x)=
\begin{cases}
1, & v_g(x)>v/2,\\
\displaystyle
\frac{1}{2} + \frac{v_g(x)}{v}, & -v/2\leqslant v_g(x) \leqslant v/2,\\
0, & v_g(x)<-v/2
\end{cases}
\end{equation}
with $v>0$ and $v_g(x) = v_g - x^2/2$. Here and below, we employ dimensionless variables by introducing characteristic scales $E_0=g^2/\alpha$ of energy, $l_0=\sqrt{g/\alpha}$ of length, and $\omega_0=\sqrt{g/m}$ of frequency (or time $t$) from a comparison of the quartic vibron potential in $H_{\rm vib}$ and the electron-vibron coupling $H_c$. Specifically, we introduce the reduced variables  $x=X/l_0$, $p=P/m\omega_0 l_0$, $\tau=\omega_0t$, $v=V/E_0$, $v_g= V_g/E_0$, and $r=R\omega_0/E_0$. In terms of these variables, we can also write $H_\textrm{vib}+H_c = E_0[{p^2}/{2} +(-\epsilon+\hat n) {x^2}/{2} + {x^4}/{4}]$ in terms of a reduced compressional force $\epsilon=-m\omega^2/g$.

The current-induced force $-x n_0(x)$ has dramatic effects on the Euler
instability, as follows from a stability analysis of the vibrational motion. The
(meta)stable positions of the nanobeam are obtained by setting the effective
force $f_{\rm eff}(x) = \epsilon x - x^3 -x n_0(x)$ to zero. Our results are
summarized in the stability diagrams in Fig.~\ref{fig:stable_solutions}. The most striking results of this analysis are: (i) The current flow renormalizes the critical force required for buckling towards larger values. (ii) At low biases, the buckled state can appear via a discontinuous transition. 

These results can be understood most directly in terms of the potential $v_{\rm eff}(x)$ associated with $f_{\rm eff}(x)$. Focusing on the current-carrying region (shown in grey in Fig.\ \ref{fig:stable_solutions} and delineated by ${\rm max}\{0,\epsilon_-\}<x^2<\epsilon_+$ with $\epsilon_\pm = 2v_g\pm v$), we find
\begin{equation}
v_{\rm eff}(x) = \frac{1}{2} \left(-\epsilon + \frac{v+2v_g}{2v} \right)x^2 +\frac{1}{4}\left(1-\frac{1}{2v}\right) x^4.
\label{effpot}
\end{equation}
The quadratic term shows that the current indeed stabilizes the unbuckled state,
renormalizing the critical force to $\tilde\epsilon=1/2+v_g/v$ when
$\epsilon_-<0<\epsilon_+$ (Fig.~\ref{fig:stable_solutions}a-b). Remarkably,
however, the current-induced contribution to the quartic term is negative at
small $x^2$ and thus {\em destabilizes} the unbuckled state. According to
Eq.~(\ref{effpot}), the quartic term in the current-induced potential becomes
increasingly significant as the bias voltage $v$ decreases and we find that the
overall prefactor of the quartic term becomes {\em negative} when $v<1/2$.
\cite{footnote_singularity} It is important to note that this does not imply a
globally unstable potential since the current-induced force contributes only for
small $x^2$. A sign reversal of the quartic term is also a familiar occurrence
in the Landau theory of tricritical points which connect between second- and
first-order transition lines. \cite{chaikin} In close analogy, the sign reversal of the quartic term in the effective potential (\ref{effpot}) signals a discontinuous Euler instability which reverts to a continuous transition at biases $v>1/2$ where the prefactor of the quartic term remains positive. 

Specifically, when $v>1/2$ (Fig.\ \ref{fig:stable_solutions}b,d), the
current-induced potential renormalizes the parameters of the vibronic
Hamiltonian but leaves the quartic term positive. This modifies how the position
of the minimum depends on the applied force in the conducting region ${\rm
max}\{0,\epsilon_-\}<x^2<\epsilon_+$, but the Euler instability remains
continuous. When $v<1/2$, the equilibrium position at finite $x$ becomes
unstable within the entire current-carrying region. This leads to a
discontinuous Euler transition when $\epsilon_-<0<\epsilon_+$ (Fig.\
\ref{fig:stable_solutions}a) and to multistability in the region $\epsilon_- <
x^2 < \epsilon_+$ when $\epsilon_- > 0$ (Fig.\ \ref{fig:stable_solutions}c). \cite{Nori}

\begin{figure}[b]
\includegraphics[width=\columnwidth]{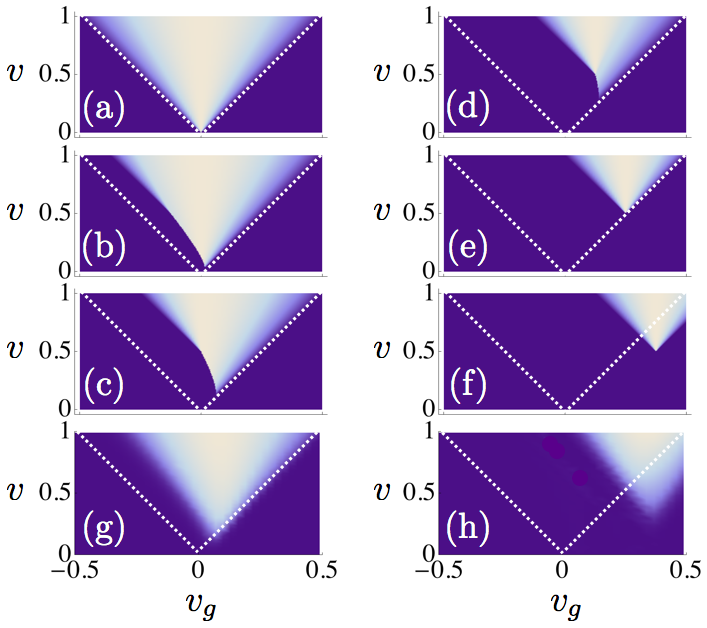}
\caption{\label{fig:conductance}%
(color online)  Conductance $G=RI/V$ in the $v_g$--$v$ plane
for applied force
(a) $\epsilon\leqslant0$,
(b) $\epsilon=0.25$,
(c,g) $\epsilon=0.5$,
(d) $\epsilon=0.75$,
(e) $\epsilon=1$,
(f,h) $\epsilon=1.25$,
within (a-f) stability analysis and (g-h) full Langevin dynamics ($r=\gamma_e=T=0.01$). Color scale: $G=0\rightarrow 1/4$ from dark blue to white. Dotted lines delineate the Coulomb diamond for $g=0$.}
\end{figure}

At the level of the stability analysis, we can also obtain the current $I$ by evaluating the rate-equation result \cite{dittrich} 
$RI(x)/V=1/4-[v_g(x)/v]^2$ at the position of the most stable minimum. Corresponding results in the $v_g$--$v$ plane are shown in Fig.~\ref{fig:conductance}a-f for various values of the applied force $\epsilon$. 
By comparison with the Coulomb diamond in the absence of the electron-vibron
coupling (dotted lines in Fig.\ \ref{fig:conductance}), we see that the Euler
instability leads to a current blockade over a significant parameter range. For
$v>1/2$, the blockade is a manifestation of the Franck-Condon blockade, \cite{koch05_PRL, pisto07_PRB} caused by the induced linear electron-vibron coupling when expanding Eq.\ (\ref{eq:H_c}) about the buckled state.

In contrast, for $v<1/2$, the current blockade is a direct consequence of the discontinuous Euler instability. We have seen above that in this regime, the buckled state becomes unstable throughout the entire current-carrying region. As a result, the current-induced force will always drive the system out of the current-carrying region, explaining the current blockade. An intriguing feature of this novel tricritical current blockade is the curved boundary of the apparent Coulomb-blockade diamond (Fig.\ \ref{fig:conductance}), a behavior which is actually observed in nanoelectromechanical systems. 

\section{Langevin dynamics}
To investigate the robustness of the stability analysis against fluctuations, we turn to the complete vibronic Langevin dynamics $\ddot x+\gamma(x)\dot x=f_\textrm{eff}(x)+\xi(\tau)$. The fluctuating force $\xi(\tau)$ is generated by fluctuations of the electronic occupation and the frictional force $-\gamma(x)\dot x$ by the delayed response of the electrons to the vibronic dynamics. To compute $\gamma(x)$ and $\xi(\tau)$, we solve Eq.\ (\ref{BL}) including the vibronic dynamics and the Langevin sources. Writing separate equations for average and fluctuations of the occupation by setting $n = \bar{n} + \delta n$, we see that the leading correction to $\bar{n}$ arises from the Poisson bracket, yielding $\bar {n} = n_0 - \frac{1}{\Gamma_++\Gamma_-}  \dot X \partial_X n_0$. At the same time, the fluctuations $\delta n$ obey the correlator $\langle \delta n(t)\delta n(t')\rangle = \frac{2}{\Gamma_++\Gamma_-} n_0 (1 -n_0) \delta(t-t')$. Inserting these results into the expression for the current-induced force $-gX n$ and employing reduced units, we find $\langle \xi(\tau)\xi(\tau')\rangle = D(x)\delta (\tau - \tau')$ with diffusion and damping coefficients $D(x)=2rx^2n_0(1-n_0)/v$ and $\gamma(x)=-rx\partial_x n_0/v$, respectively. Finally, we can pass from the Langevin to the equivalent Fokker-Planck equation \cite{blant04_PRL,marti06_PRB} for the probability $\mathcal{P}(x, p, \tau)$ that the nanobeam is at position $x$ and momentum $p$ at time $\tau$, 
\begin{equation}
\label{eq:FP}
\partial_\tau\mathcal{P}=-p\partial_x\mathcal{P}
-f_\textrm{eff}\partial_p\mathcal{P}+\gamma\partial_p(p\mathcal{P})
+\frac{D}{2}\partial_p^2\mathcal{P}.
\end{equation}
Note that the diffusion and damping coefficients are non-vanishing only in the
conducting region. \cite{extrinsic_damping}

The current $I=\int dxdp\,\mathcal{P}_\textrm{st}(x, p)I(x)$ is now obtained from the stationary solution $\partial_\tau\mathcal{P}_\textrm{st}=0$ of the Fokker-Planck equation \eqref{eq:FP}. Numerical results for the scaled linear conductance $G=RI/V$ are shown in Fig.~\ref{fig:conductance}g-h, using the same parameters as in Fig.~\ref{fig:conductance}c,f. We observe that the fluctuations reduce the size of the blockaded region and blur the edges of the conducting regions as the system can explore more conducting states in phase space. Nevertheless, the conclusions of the stability analysis clearly remain valid qualitatively.  

\section{Conclusion}
We have presented a general approach to the interplay
between continuous mechanical instabilities and current flow in
nanoelectromechanical systems, and have applied our general framework to the
Euler buckling instability. The current flow modifies the nature of the buckling
instability from a continuous to a tricritical transition. Likewise, the
instability induces a novel tricritical current blockade at low bias. Our
nonequilibrium Born-Oppenheimer approach generalizes not only to other
continuous mechanical instabilities, but also to other systems such as
semiconductor quantum dots or single-molecule junctions with a discrete
electronic spectrum, to other types of electron-vibron coupling, \cite{footnote_other} and to further transport characteristics (e.g., current noise). 

Our proposed setup can be realized experimentally by clamping, e.g., a suspended carbon nanotube and applying a force to atomic precision either using a break junction or an atomic force microscope. Indeed, several recent experiments 
show that the electron-vibron coupling is surprisingly strong in suspended
carbon nanotube quantum dots. \cite{letur09_NaturePhysics,huettel_Science09,bachtold_Science09}
 
\begin{acknowledgments}
We acknowledge financial support through Sfb 658 of
the DFG
(GW, EM, FvO) and ANR contract JCJC-036 NEMESIS (FP). FvO enjoyed the hospitality of the KITP (NSF PHY05-51164).
\end{acknowledgments}


\begin{thebibliography}{}
\bibitem{footnote_zeromode}
The rotation of the plane of the buckled nanobeam is assumed massive due to clamped boundary conditions.
\bibitem{footnote_singularity} We note that the singularity at small $v$ is cut off for bias voltages of the order of temperature or level broadening.
\end{thebibliography}
\end{document}